\documentstyle[11pt,newpasp,twoside,epsf]{article}
\def\edcomment#1{\iffalse\marginpar{\raggedright\sl#1\/}\else\relax\fi}
\reversemarginpar

\begin{document}
\title{
Chemical Evolution of the Stellar and Gaseous Components of Galaxies
in Hydrodynamical Cosmological Simulations}
 
\author{Cora, S. A.}
\affil{Observatorio Astron\'omico de La Plata, 
Paseo del Bosque s/n, 1900 La Plata, Argentina.}
\author{Tissera, P. B.}
\affil{
Instituto de Astronom\'{\i}a y F\'{\i}sica del Espacio,
Buenos Aires, Argentina.}
\author{Lambas D. G. and 
Mosconi, M. B.}
\affil{
Observatorio Astron\'omico de C\'ordoba, Argentina.}

\begin{abstract}
We present preliminary results on the effects of mergers
on the chemical properties of galactic objects in hierarchical
clustering  scenarios. We adopt a hydrodynamical chemical
code that allows to describe the coupled evolution of
dark matter and baryons within a cosmological context. 
We found that disk-like and spheroid-like objects have
distinctive metallicity patterns that may be the result
of different evolution. 
\end{abstract}


The implementation of a chemical model in a cosmological code
provides a useful tool for the study of the chemical properties of galaxies
in relation to their formation and evolution in a cosmological framework. 
We performed numerical simulations with 
the hydrodynamical AP3MSPH code 
that includes
star formation (SF) and metal enrichment from nucleosynthesis
ejecta of supernovae type I and type II (Mosconi et al. 2000).
This code allows the description of the star
formation and chemical history of the gas and stars 
of galactic objects in hierarchical clustering scenarios
where mergers and interactions play a crucial role in 
regulating SF.

The chemical content of the stellar population  
and the interstellar medium (ISM)
of the identified galaxy-like objects (GLOs) at $z=0$ is determined by the way 
in which the SF proceeds in the different clumps that assemble
to form the final object. We found GLOs at $z=0$ that have
stellar components  with 
mean age metallicity relations 
and abundance distribution functions, 
such as the 
relation between [O/Fe] and [Fe/H], that 
reproduce fairly well the observed patterns for the Milky-Way.

The chemical abundances in the ISM in gas-rich galaxies
allow to trace the evolution of individual galaxies.
The steep negative abundance gradients of the gaseous component found for the GLOs
are within the expected values for the oxygen and nitrogen, 
as result from the comparison with
HII regions and early B-type main sequence objects. 
We found that spheroid-like objects have slopes 
less pronounced 
than those GLOs where the gas forms a well-defined disk,
indicating very different histories of formation and evolution where 
the mergers have a significant role.
In order to understand the way in which mergers affect the 
chemical abundances in the 
ISM, 
we followed the
chemical evolution of the gaseous
component in a typical GLO in one of the simulations
during its main merger event.
Figure 1a shows the evolution with redshift of
the gas mean oxygen abundances as a function of the mass-fraction in 
concentric shells, while in figure 1b we have the SF rate of this
GLO with the merger event clearly pointed out.  
The metallicity profile is steep before the merger. 
As the companion approaches, it raises up because of the 
enhancement of chemical production 
at the central regions, produced by
the tidally induced 
starburst (Tissera 2000).
The final profile is flatter than the original one, 
reflecting the fact that 
gas particles of different metallicities have been well-mixed
during the merger. A detailed study of the effects
of mergers on the matter distribution is on preparation 
(Cora et al. 2001).

\begin{figure}
\plotfiddle{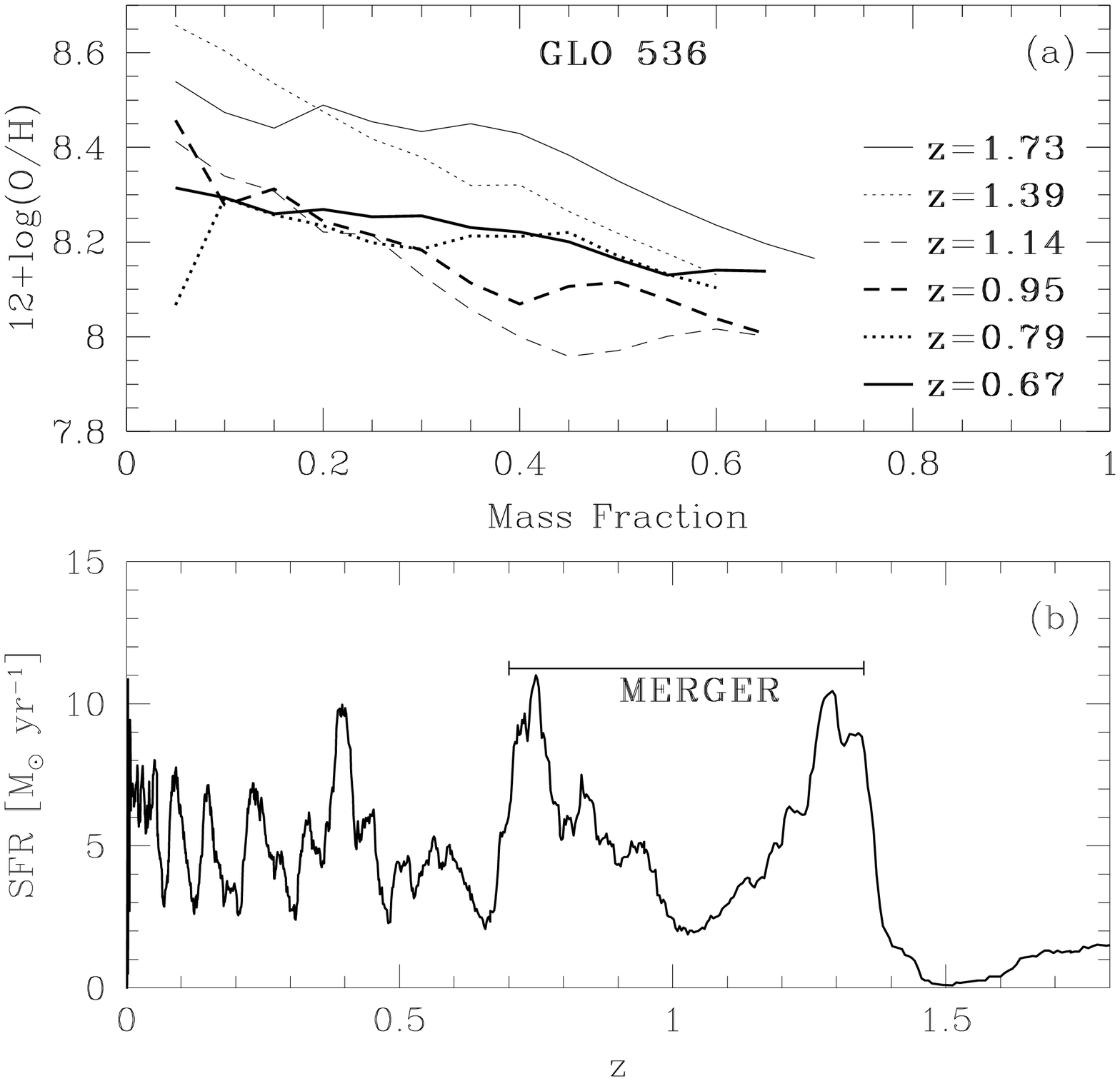}{2.3in}{0}{65}{35}{-190}{-65}
\caption{
(a) Mean oxygen gas abundances as a function of the mass
fraction for a typical GLO in one of the simulations at six different redshifts.
(b) SFR for the same GLO. The merger event is pointed out.}
\end{figure}

To sum up, in our simulations,
GLOs have different evolutionary histories in consistency with a hierarchical
clustering scenario that affect their SF rates and chemical evolution.
This chemical model can take all these physical processes into account, 
resulting in a powerful tool for studying galaxy formation.


\end{document}